\documentclass[12pt,notitlepage]{article}

\usepackage{amsmath}         
\usepackage{amssymb}         
\usepackage{amsfonts}        
\usepackage{graphicx}        

\usepackage{color}         
\usepackage{slashed}       
\usepackage{framed}        
\usepackage{subcaption}    
\usepackage[mathscr]{euscript} 
\usepackage{cite}          
\usepackage{booktabs}      
\usepackage{youngtab}	    
\usepackage{arydshln} 	    
\usepackage{tocloft}		   
\usepackage{setspace}	   
\usepackage{listings}       
\usepackage[margin=2cm]{geometry}   
\graphicspath{{figures/}}	        
\numberwithin{equation}{section}    
\renewcommand{\tilde}{\widetilde}   

\newcommand{\email}[1]{\href{mailto:#1}{#1}}

\newenvironment{institutions}[1][2em]{\begin{list}{}{\setlength\leftmargin{#1}\setlength\rightmargin{#1}}\item[]}{\end{list}}

\let\oldenumerate\enumerate
\renewcommand{\enumerate}{
  \oldenumerate
  \setlength{\itemsep}{1pt}
  \setlength{\parskip}{0pt}
  \setlength{\parsep}{0pt}
}

\let\olditemize\itemize
\renewcommand{\itemize}{
  \olditemize
  \setlength{\itemsep}{1pt}
  \setlength{\parskip}{0pt}
  \setlength{\parsep}{0pt}
}

\usepackage{bbm}

\usepackage[
	colorlinks=true,
	citecolor=black,
	linkcolor=black,
	urlcolor=blue,
	hypertexnames=false]{hyperref}

\newcommand{\gev}{\ensuremath{\text{\small GeV}}}

\newcommand{\abs}[1]{\ensuremath{\left|#1\right|}}
\newcommand{\ZZ}{\ensuremath{\mathbbm{Z} }}

\newcommand{\Pf}{\ensuremath{\text{Pf}\,}}

\newcommand{\barA}{\ensuremath{\overline{A}}}

\newcommand{\barQ}{\ensuremath{\overline{Q}}}

\newcommand{\q}{\ensuremath{q}}
\newcommand{\barq}{\ensuremath{\overline{q}}}
\newcommand{\p}{\ensuremath{\overline{q}}}

\newcommand{\upq}{\ensuremath{U(1)_\text{PQ}}}

\newcommand{\tJ}{\ensuremath{\tilde{J}}}
\newcommand{\tK}{\ensuremath{\tilde{K}}}
\newcommand{\tM}{\ensuremath{\tilde{M}}}

\newcommand{\tX}{\ensuremath{\tilde{X}}}
\newcommand{\tY}{\ensuremath{\tilde{Y}}}
\newcommand{\tZ}{\ensuremath{\tilde{Z}}}
\renewcommand{\P}{\barQ}
\newcommand{\arr}{\ensuremath{r}}
\newcommand{\bG}{\ensuremath{\tilde{G}}}
\newcommand{\tG}{\ensuremath{\tilde{G}}}

\newcommand{\x}{\ensuremath{x}}
\newcommand{\y}{\ensuremath{y}}
\newcommand{\z}{\ensuremath{z}}
\newcommand{\sq}{\ensuremath{\psi}}
\newcommand{\tx}{\ensuremath{\tilde{\x}}}
\newcommand{\ty}{\ensuremath{\tilde{\y}}}
\newcommand{\tz}{\ensuremath{\tilde{\z}}}

\newcommand{\yI}{\ensuremath{{\bf 1}}}
\newcommand{\yF}{\ensuremath{\tiny\yng(1)}}
\newcommand{\yFb}{\ensuremath{\tiny \overline{\yng(1)}}}
\newcommand{\yA}{\ensuremath{\tiny\Yvcentermath1 \yng(1,1)}}
\newcommand{\yAb}{\ensuremath{ \tiny\Yvcentermath1\overline{\yng(1,1)}}}
\newcommand{\yAd}{\ensuremath{{\bf Adj}}}
\newcommand{\Lb}[1]{\ensuremath{ \Lambda_{#1}^b}}
\newcommand{\tL}{\ensuremath{ \tilde{\Lambda}}}
\newcommand{\tLb}[1]{\ensuremath{ \tilde{\Lambda}_{#1}^b}}

\newcommand{\mpl}{\ensuremath{M_\text{P}}}
\newcommand{\ev}[1]{\ensuremath{ \langle #1 \rangle} }
\renewcommand{\eqref}[1]{Eq.~(\ref{#1})}

\newcommand{\susy}{\ensuremath{\textsc{susy}}}

\newcommand{\wtree}{\ensuremath{W_\text{tree}}}
\newcommand{\pngb}{pNGB}
\newcommand{\ngb}{NGB}
\newcommand{\thmax}{\ensuremath{\abs{\theta_\text{max}}}}

\newcommand\Tstrut{\rule{0pt}{2.5ex}}         
\newcommand\Bstrut{\rule[-1.3ex]{0pt}{0pt}}

\usepackage{xcolor}

\begin{document}

\begin{flushright}
UCI-HEP-TR-2017-07
\end{flushright}

\begin{center}

    {\LARGE \bf 
   A Composite Axion\\ from a Supersymmetric Product Group}

    \vskip 1cm

    { \bf Benjamin~Lillard} and {\bf Tim~M.P.~Tait} 
    \\ \vspace{-.2em}
    { 
    \footnotesize
    \email{blillard@uci.edu},~
    \email{ttait@uci.edu}
    }
	
    \vspace{-.2cm}

    \begin{institutions}[2.25cm]
    \footnotesize
    {\it 
	    Department of Physics and Astronomy, University of California, Irvine, CA 92697, USA
	    }   
    \end{institutions}

\end{center}
\vspace*{0.5cm}

\begin{abstract}
 \noindent
A global $\upq$ symmetry is protected from gravitational effects in the s-confining $SU(N)^k$ product group theory with $A+4Q +N\barQ$ matter.
If the $SU(4)$ family symmetry is gauged and an appropriate tree-level superpotential is added, then the dynamically generated superpotential spontaneously breaks $SU(4)\times \upq \rightarrow SU(3)_c$ and produces a QCD axion. Small values of the $CP$-violating $\theta$ parameter are then possible without any fine-tuning, as long as the product group is suitably large.
By introducing a second copy of the s-confining $SU(N)$ product group also coupled to the gauged $SU(4)$, we find that values as small as 
$N=7$ are consistent with $\bar\theta<10^{-10}$, even under the pessimistic assumption that the dominant contribution to the axion quality is at
tree level.

 \end{abstract}

\section{Introduction}

Despite its success at predicting the results of particle experiments, the Standard Model remains widely unloved. Its unpopularity is due in part to a few inexplicably small parameters, including the $\mathcal O(10^{-16})$ ratio between the electroweak and Planck scales, the puzzling array of Yukawa couplings,
and the degree to which QCD conserves the discrete charge ($C$) and parity ($P$) symmetries, $\abs{\theta} < 10^{-10}$.
In addition, the Standard Model is clearly incomplete, failing to describe gravitation, dark matter, and neutrino masses.

Prominent solutions to these theoretical shortcomings include supersymmetry (\susy ), 
which stabilizes the electroweak scale and can support dark matter; 
extra dimensions and composite models, which can generate hierarchies dynamically; 
and axions, which explain the smallness of the QCD $CP$ parameter $\theta$ while supplying a dark matter candidate.
In this paper we consider a hybrid of these elements, a supersymmetric composite axion model, as a solution to the strong $CP$ problem that is free from fine-tuning.


At issue (for more complete discussion, see Refs.~\cite{Peccei:1996ax,Kim:2008hd}) is
the $\theta$ term of the QCD Lagrangian,
\begin{equation}
\mathcal L = \frac{g^2}{32\pi^2} \bar\theta ~\epsilon^{\mu\nu\rho\sigma} G_{\mu \nu}^a G_{\rho\sigma}^a  
\equiv \frac{g^2}{32\pi^2} \bar\theta ~G_{\mu \nu}^a \tilde{G}^{a\mu\nu},
\end{equation}
which violates both $P$ and $CP$.
$\bar\theta$ is the physical combination of the intrinsic coefficient $\theta$ and a phase in the quark mass matrix,
\begin{equation}
\bar\theta \equiv \theta + \text{arg}\, \text{det}\, M_Q.
\end{equation} 
Measurements of the neutron electric dipole moment require $\abs{\bar\theta}<10^{-10}$ \cite{Afach:2015sja}.  Such a tiny value
appears to require an extraordinary cancellation between two apparently unrelated quantities.


In a simple axion model, $\bar\theta$ is associated with the transformation parameter of an approximate global $\upq$ symmetry 
\cite{Peccei:1977ur,Peccei:1977hh,Wilczek:1977pj,Weinberg:1977ma,Kim:1979if,Shifman:1979if,Dine:1981rt}. 
\upq\ is spontaneously broken at some high scale $f_a$ by the expectation value of a 
\upq-charged scalar field or the formation of a \upq-charged fermion condensate, resulting in a pseudo-Nambu--Goldstone boson (\pngb):
the axion $a$.
Due to the nonzero $SU(3)_c^2$-$\upq$ anomaly, non-perturbative QCD dynamics induce an expectation value for the axion
such that $CP$ is a symmetry of the vacuum, and the axion acquires a small mass.

At energies below $f_a$, the effective Lagrangian contains the term:
\begin{equation}
\mathcal L = \frac{g^2}{32\pi^2} \left( \bar\theta + \mathcal A \frac{a}{f_a} \right) G_{\mu \nu}^a \tilde{G}^{a\mu\nu},
\end{equation}
where $\mathcal A$ is the $SU(3)_c^2$-$\upq$ anomaly coefficient.
Nonperturbative QCD generates a periodic potential for the axion which can be heuristically
described by
\begin{equation}
V[a] = m_\pi^2 f_\pi^2 \left( 1 - \cos\left[ \mathcal A \frac{a}{f_a} + \bar\theta\right] \right), \label{eq:vainst}
\end{equation}
where $m_\pi$ and $f_\pi$ are the pion mass and decay constant, respectively.
This potential is minimized when $\ev{a} = -f_a \bar\theta /\mathcal A$, leading to
$CP$ conservation in the vacuum. We choose to normalize the \upq\ charges so that $\mathcal A=1$,
for which the axion mass is\footnote{
More careful treatments based on the QCD chiral Lagrangian~\cite{DiVecchia:1980yfw} result in a potential given by:
$V[a]= m_\pi^2 f_\pi^2 \left( 2 - \sqrt{1+ \frac{2 m_u m_d}{(m_u + m_d)^2 } \left( \cos\left[\mathcal A \frac{a}{f_a} +\bar\theta\right] \right)   } \right)$, where
$m_{u,d}$ are the up- and down-quark masses, and leading to an axion mass
$m_a^2 = \frac{m_u m_d}{(m_u+m_d)^2} \frac{m_\pi^2 f_\pi^2}{f_a^2}$.  The distinction between these two expressions for $V[a]$ is unimportant in
terms of assessing the axion quality, and we use Eq.~\ref{eq:vainst} for our analysis.},
\begin{equation}
m_a^2 = \frac{m_\pi^2 f_\pi^2}{f_a^2}.  
\label{eq:vamasses}
\end{equation}
Experimental observations set bounds on the value of $f_a$. A lower bound $f_a \gtrsim 10^{9} ~\gev $ is 
derived from constraints on stellar and supernova cooling~\cite{Raffelt:2006cw}, while the axion relic abundance 
suggests $f_a \lesssim10^{12}\, \gev$ in the absence of cosmological fine tuning~\cite{Fox:2004kb}.


\paragraph{Axion Quality Problem:}
Simple axion models are plagued by the theoretical inconsistencies endemic to theories containing fundamental scalar fields. The expectation 
value of the new complex scalar $\ev{\phi} \sim f_a$ receives additive corrections from high-energy physics which,
while less severe than the electroweak hierarchy \cite{deGouvea:2014xba}, remains a concerning source of fine-tuning.
Models of axions also suffer from a different concern which is potentially much more troubling: 
the \emph{axion quality problem}. Any \upq-violating effects in the scalar potential can shift the axion VEV away from $\theta=0$,
inducing the strong $CP$ problem rather than solving it. 
In particular, non-perturbative quantum gravity is expected to violate global symmetries
\cite{Giddings:1987cg,Lee:1988ge,Kamionkowski:1992mf,Barr:1992qq,Kallosh:1995hi,Alonso:2017avz}, 
leading to terms in the low energy effective action of the form
\begin{equation}
\mathcal L_g \sim \frac{\abs{\phi}^p (\phi + \phi^\star) }{\mpl^{p-3}}, \label{eq:lgex}
\end{equation} 
which is inconsistent with $\abs{\theta}<10^{-10}$ unless the $p=4$ term has a coefficient smaller than $\mathcal O(10^{-55})$. 
Considering that the axion is introduced to explain fine-tuning of $\mathcal O(10^{-10})$, this calls its motivation into serious question,
and any successful axion model must prevent linear shifts of the form $\ev{a} \rightarrow \ev{a} + f_a \Delta \theta$ with $\Delta \theta > 10^{-10}$. 

More generally, we can analyze arbitrary \upq\ violation by including it in the axion potential $V[a]$ as 
\begin{equation}
\delta V[a] = (Q~ f_a^4 ) \cos \left(\kappa \left[\frac{a}{f_a} + \bar\theta\right] + \theta_0\right), \label{eq:quality}
\end{equation}
for a dimensionless ``quality factor" $Q$, an integer $\kappa$ and an angle 
$\theta_0$. Experimental measurements of $\ev{\theta}$ set a maximum bound on $Q$; we derive the general expression in 
Appendix~\ref{appx:quality}. For $\kappa\sin\theta_0 \sim \mathcal O(1)$, $\abs{{\theta}}<10^{-10}$ requires:
\begin{equation}
Q < 10^{-62} \left( \frac{10^{12} ~\gev}{f_a} \right)^4 = 10^{-50} \left( \frac{10^{9}~ \gev}{f_a} \right)^4.
\end{equation}

\paragraph{Consistent Axion Models:}
Several solutions to the axion quality problem are known, in which the \upq\ is protected by associating it with new gauged symmetries. In the simplest solutions a 
gauged discrete $\ZZ_N$ symmetry~\cite{Chun:1992bn} forbids \upq-violating operators of dimensions smaller than $N$. More sophisticated models can employ 
discrete groups as small as $\ZZ_4$ while forbidding the problematic operators~\cite{Carpenter:2009zs,Harigaya:2013vja}.
Solutions without gauged discrete symmetries also exist: for example, a composite model~\cite{Randall:1992ut} with a gauged 
$SU(N)\times SU(m) \times SU(3)_c$ protects \upq\ to arbitrarily high order. More recently~\cite{DiLuzio:2017tjx}, a qualitatively different 
$SU(N)_L \times SU(N)_R \times SU(3)_c$ model has been shown to suppress Planck scale corrections appropriately.

Other constructions protect $\upq$ by gauging a related Abelian group. In one model~\cite{Cheng:2001ys} 
with a compact extra dimension, a gauged $U(1)$ symmetry is spontaneously broken by fields localized on two separated four-dimensional branes. 
One combination of the fields is eaten by the gauge field, while the other acts as the 
QCD axion and is protected from gravitational corrections. A related model~\cite{Hill:2002kq} gauges a product group of the form $U(1)^k$ with $k\geq14$, which can 
also be interpreted as a $k$ site deconstruction of a compact fifth dimension.
In a different class of models~\cite{Barr:1992qq,Fukuda:2017ylt}, the fields are assigned large and relatively prime $U(1)$ charges, so that an accidental $\upq$ is 
protected from low-dimensional operators.

Some of these models, while successful at forbidding low-dimensional \upq-breaking operators, still suffer from a hierarchy problem.  
One resolution is supersymmetry (\susy), 
which protects $f_a$ from loop-level corrections, so that the theory is technically natural if the \susy-breaking scale is not 
much larger than $f_a$. Another compelling direction is
composite models, which can suppress dangerous gravitational contributions to the axion potential while additionally offering the potential
to determine the scale of \upq\ breaking from the confining dynamics. For asymptotically free gauge theories the confinement scale is expected to be exponentially 
suppressed compared to \mpl, so the hierarchy between $f_a$ and \mpl\ can be naturally generated dynamically.

In this article, we present a qualitatively new supersymmetric composite axion model which tames both the quality and hierarchy problems. 
The axion is a composite formed of large product of fundamental fields, such that the quality problem is ameliorated by a 
sufficiently large power of $(\Lambda / \mpl)^n$, where $f_a \sim \Lambda$ 
is dynamically generated by the confinement of a product of non-Abelian gauge theories.
Supersymmetry allows for control over the low energy physics of the non-perturbative confining dynamics, and additionally stabilizes any other
mass scales (including, perhaps, the electroweak scale).
Our work is laid out as follows: in Section~\ref{sec:main}, we explore a minimal construction in terms of its UV degrees of freedom.
In Section~\ref{sec:confinement}, we analyze its low energy behavior after confinement, with 
Section~\ref{sec:ssb} discussing the breaking of the global symmetries, including \upq.  
Section~\ref{sec:grav} estimates the size of the leading gravitational corrections, and determines parameters such
that the axion quality problem is ameliorated to a sufficient degree.  In Section~\ref{sec:dynsp}, we show how a simple extension of the basic model
can dynamically generate superpotential terms on which the basic module relies, resulting in a theory in which all of the essential mass scales
are dynamically generated.  In Section~\ref{sec:conclusions}, we conclude.
As we shall see, solving the quality problem can imply that a theory whose
low energy limit looks like a rather standard invisible axion model may blossom at high energies into a rich interlocking structure of gauge dynamics.

\section{Axion from a Supersymmetric Product Group} \label{sec:main}

We consider theories in which the axion emerges as a composite in the low energy description of confining supersymmetric gauge dynamics.
In order to generate the scale $f_a$ dynamically as a by-product of confinement, we further specialize to s-confining theories~\cite{Csaki:1996sm,Csaki:1996zb},
in which a set of gauge-invariant operators provides a smooth description of the moduli space (valid at the origin), 
and a dynamically generated superpotential enforces the classical constraints.
Our basic building blocks are $SU(N)$ gauge theories with one antisymmetric $A$, four fundamental quarks $Q$, and $N$ antifundamental antiquarks $\barQ$; 
and $Sp(2n)$ gauge theories with $(2n+4)$ quarks $Q$. 
Both of these theories have been shown to s-confine~\cite{Berkooz:1995km,Poppitz:1995fh,Pouliot:1995me,Intriligator:1995ne},
and the $A+4Q+N\barQ$ module has an $SU(4)$ flavor symmetry (acting on the $Q$ fields) into which $SU(3)_c$ QCD can be embedded.

Gauging the $SU(4)$ flavor symmetry requires an additional four quarks $q$ transforming in the 
antifundamental representation of $SU(4)$ to cancel the $SU(4)^3$ anomaly.  Supplemented by an appropriately chosen external
superpotential, the $SU(N)$ confines and an appropriate $\upq$ can be spontaneously broken.  However, the resulting axion quality 
from this simple module is far from sufficient to accommodate $\abs{{\theta}}<10^{-10}$.

\begin{figure}[t] \centering
\includegraphics[scale=1.0]{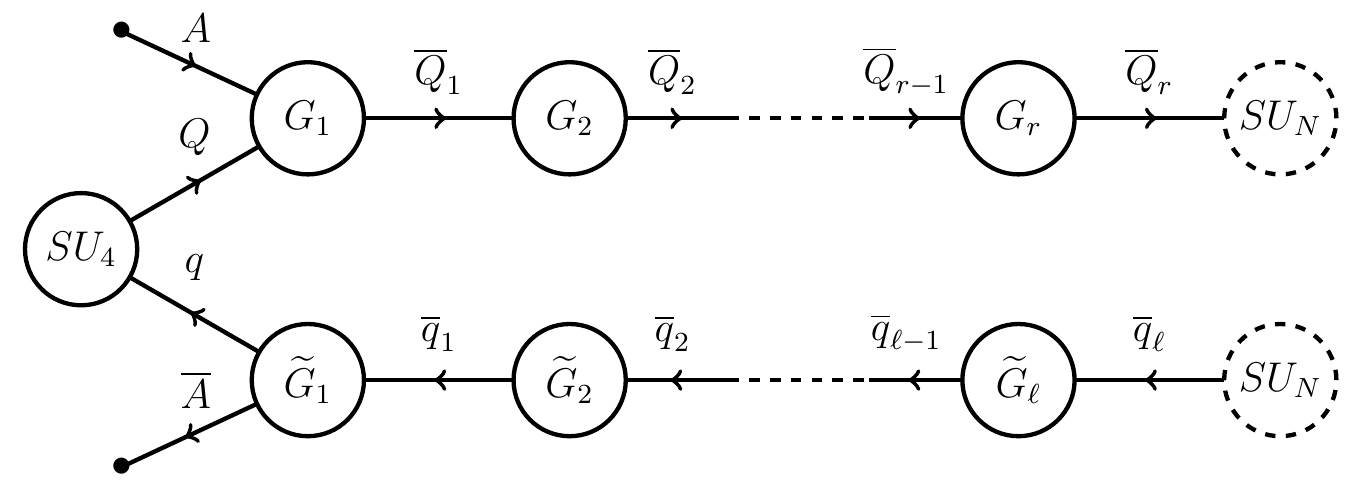}
\caption{Moose diagram indicating the matter content and gauge interactions of the $SU(N)^\ell\times SU(4)\times SU(N)^\arr$ composite axion model. 
Each $G_i$ and $\bG_i$ corresponds to a gauged $SU(N)$, whereas $SU(N)$ flavor symmetries are represented by dashed circles. 
The bifundamental fields $Q$, $\barQ_i$, $\q$, and $\barq_i$ are depicted as directed line segments connecting adjacent groups, while the field 
$A$ ($\barA$) transforms under $G_1$ ($\bG_1$) in the antisymmetric two-tensor representation.
 }
\label{fig:moose}
\end{figure}

High axion quality can be enforced by expanding the $SU(N)$ into a product group.
It has recently been demonstrated that s-confining product group models can be constructed by gauging the $SU(N)$ flavor 
symmetry of the $A+4Q+N\barQ_1$ theory, such that the field $\P_1$ transforms as a bifundamental under $SU(N)\times SU(N)$, 
with $N$ quarks $\P_2$ canceling the anomalies~\cite{Lillard:2017mon}. 
Iterating to $SU(N)^k$, the matter fields include the $SU(N)_{(1)}$-charged $A+4Q$; 
a string of $SU(N)_{(i)} \times SU(N)_{(i+1)}$ bifundamentals $\barQ_i$; 
and $N$ fields $\barQ_k$ charged only under the gauged $SU(N)_{(k)}$.
The gauge-invariant operators include ``mesons" of the form $(Q \P_1 \P_2 \ldots \P_k)$ and $(A \P_1^2 \ldots \P_k^2)$; ``baryons" $(\P_i^N)$ for each $i=1\ldots k$; 
and special baryons $(A^{\frac{N-p}{2}} Q^{p})$ for $0\leq p\leq 4$, subject to the condition that $(N-p)$ is even.
An axion living in a combination of these fields enjoys the feature that 
increasing $k$ and $N$ results in increasingly suppressed gravitational corrections.

Extending the gauge symmetries on both sides, we arrive at a theory in which the 
full matter content is $\{A, Q, \P_1 \ldots \P_\arr; \barA, q, \p_1 \ldots \p_\ell\}$, with the gauge group $SU(N)^\ell \times SU(4) \times SU(N)^\arr$. 
The gauge structure and matter assignments is represented as a moose diagram in Figure~\ref{fig:moose}, and is vaguely reminiscent of
a deconstructed extra dimension with a bulk $SU(N)$ broken to $SU(4)$ on a defect.
For convenience, we introduce the notation $SU(N)^\ell = \tG_1 \times \tG_2 \times \ldots \times \tG_\ell$ and $SU(N)^\arr = G_1 \times G_2 \times \ldots \times G_\arr$, 
where $\tG_i$ and $G_i$ confine at scales $\tL_i$ and $\Lambda_i$ respectively.
Up to a constant, the holomorphic scales $\tL_i$ and $\Lambda_i$ are defined as
\begin{align}
\tLb{i} \equiv \mu^b \exp\{ -8\pi^2/\tilde{g}_i^2 + i \tilde{\theta}_{i} \} &,&
\Lb{i} \equiv \mu^b \exp\{ -8\pi^2/g_i^2 + i \theta_{i} \} ,
\end{align}
where $\tilde{g}_i$ and $g_i$ are the coupling constants of the gauge groups $\tG_i$ and $G_i$. In the dynamically generated superpotential for each group there is an overall constant that is not determined by symmetry arguments; to simplify the notation, we absorb these constants into $\tLb{i}$ and $\Lb{i}$. 

In the absence of an external superpotential, there is a conserved $U(1)_A\times U(1)_B \times U(1)_C \times U(1)_R \times SU(N)_L \times SU(N)_R$ 
global symmetry, and an approximate $\upq$ that is  broken by the $SU(4)^2$-$U(1)$ anomaly. Charges are shown in Table~\ref{table:UVtheory}, where for convenience, we have taken the $U(1)_R$ charges of $Q$ and $A$ to be equal to $\q$ and $\barA$, respectively, with $q_Q = \frac{N-4}{N}$ and $q_A= \frac{16-2N}{N(N-2)}$.
By defining \upq\ as in Table~\ref{table:UVtheory}, we assume that the operator $(A \P_1^2 \ldots \P_\arr^2)$ is more suppressed than $(\barA \barq_1^2 \ldots \barq_\ell^2)$, 
so that \upq\ is expected to be a better symmetry than $U(1)_A$. 
Appropriate \upq\ charges in the opposite limit can be recovered by performing the following outer automorphism on the moose diagram:
\begin{align}
\ell \leftrightarrow \arr &,&
G_i \leftrightarrow \bG_i &,&
\Lambda_i \leftrightarrow \tL_i &,&
A \leftrightarrow \barA &,&
Q \leftrightarrow q &,&
\P_i \leftrightarrow \p_i .
\end{align}

\begin{table}[t]
\hspace*{-0.75cm}
\begin{tabular}{| c | c | c c c | c | c c c | c || c c c c | c |} \hline
\Tstrut \Bstrut
   &$SU(N)_L$	&$\bG_\ell$ &$\ldots$&$\bG_1$&$SU(4)$&$G_1$	&$\ldots$&$G_\arr$&$SU(N)_R$	&$U_A$	& $U_B$	& $U_C$	& $U_R$	& $\upq$	\\ \hline
$\p_\ell$  &\yF	&	\yF	&		&		&		&		& 		&		&		&	0	&	0	& $\pm1$	&	0	&	0	\\ 
$\p_{\ell-1}$&	&	\yFb	&	\yF	&		&		&		& 		&		&		& $0$	&	0	&$\mp1$	&	0	&	0	\\
$\vdots$&		& 		&$\ddots$	&		&		&		&		&		& 		&$\vdots$&$\vdots$	&$\vdots$&$\vdots$	&	$\vdots$	\\ 
$\p_1$&		& 		&	\yFb	&	\yF	&		&		&		&		&		&	0	&	0	&	1	&	0	&	0	\\ 
$\barA$&		&		&		&	\yAb	& 		&		&		&		&		&	$-4$	& 0 &$\frac{-N}{N-2}$&$q_A$	&	0	\\ 
$\q$&		& 		&		&	\yFb	&	\yFb	&		&		&		&		&	$N-2$&	0	&	0	&$q_Q$	&	0	\\ \hline
\Tstrut
$Q$&		&		&		&		& 	\yF	&	\yF	&		&		&		&$2-N$	&	0	&	0	&$q_Q$	&$\frac{2-N}{N}$ \\ 
$A$&		&		&		& 		&		&	\yA	&		&		&		&$4$	 &$\frac{-N}{N-2}$&	0 	&$q_A$	&$4/N$	\\ 
$\P_1$&		&		&		& 		&		&	\yFb	&	\yF	&		&		&	0	&	1	&	0	&	0	&	0	\\ 
$\vdots$&		&		&		& 		&		&		&$\ddots$	&		&		&$\vdots$&$\vdots$	&$\vdots$	&$\vdots$	&	$\vdots$	\\ 
$\P_{\arr-1}$&	&		&		& 		&		&		&	\yFb	&	\yF	&		&	0	&$\mp1$	&	0	&	0	&	0	\\ 
$\P_{\arr}$&	&		&		& 		&		&		&		&	\yFb	&	\yF	&	0	&$\pm1$	&	0	&	0	&	0	\\ \hline
\end{tabular}
\caption{Representations of the matter fields under the gauged $SU(N)^\ell \times SU(4) \times SU(N)^\arr$ symmetries, 
the flavor symmetries $SU(N)_L \times SU(N)_R \times U(1)^4$, and the approximate \upq\ symmetry. 
}  
\label{table:UVtheory}
\end{table}

At a generic point on the moduli space the full global symmetry is spontaneously broken, producing a number of Nambu-Goldstone bosons. 
Although the explicit symmetry breaking from gravity would supply masses for the \pngb s, a tree-level external superpotential 
\begin{equation}
\wtree = \frac{ (\barA \barq_1^2 \barq_2^2 \ldots \barq_\ell^2)}{M_A^{2\ell-2}} + \frac{(\P_1^N)}{M_B^{N-3}} + \frac{(\p_1^N)}{M_C^{N-3}} + \frac{(A^{m} Q)(A^{m-1} Q^3)}{M_R^{N-1} } + \frac{(\barA^{m} \q)(\barA^{m-1} \q^3)}{M_r^{N-1} } \label{eq:wtree}
\end{equation}
increases the \pngb\ masses by breaking the global symmetries more severely.
This is essential in the case of the second ($M_B$) term, which as we shall see below determines the 
PQ symmetry breaking scale $f_a$ after confinement.
The remaining $M_i$ could be safely taken to be $\mpl$ without harm.  In addition,
to avoid deforming the $G_1$ confinement, we choose them to satisfy $\Lambda_1\lesssim M_i$. 

In Section~\ref{sec:dynsp} we discuss the possibility that some of the terms in \eqref{eq:wtree} are generated dynamically through the s-confinement of a strongly coupled $Sp(2n)$ gauge group, providing a natural and completely dynamical origin for the scale $f_a$.

\subsection{Confinement}
\label{sec:confinement}

We choose the UV gauge couplings such that  $SU(N)^\ell$ and $SU(N)^\arr$ confine at an intermediate
scale where $SU(4)$ remains weakly coupled and supersymmetry is unbroken. 
For odd
$N=2m+1$, the groups $SU(N)^\ell$ and $SU(N)^\arr$ confine separately to produce the following 
hadrons:
\begin{align}
J_L = (\barq_\ell \barq_{\ell-1} \ldots \barq_1 \q) &,&
K_L =(\barq_\ell^2 \barq_{\ell-1}^2 \ldots \barq_1^2 \barA) &,&
\x_1 = (\barA^m \q) &,&
\y_1 = (\barA^{m-1} \q^3) &,&
\z_i = (\barq_i)^N ,\\
J_R = (Q \barQ_1 \barQ_2 \ldots \barQ_\arr) &,&
K_R =(A \barQ_1^2 \barQ_2^2 \ldots \barQ_\arr^2) &,&
X_1 = (A^m Q) &,&
Y_1 = (A^{m-1} Q^3) &,&
Z_i = (\barQ_i)^N .
\end{align}
Their transformation properties under the global symmetries are summarized in Table~\ref{table:IRtheory}.
These operators obey quantum-modified equations of motion, for which we define the shorthand notation:
\begin{align}
(\tilde{\Pi}_{1}^{\ell} \z) &= \left\{ \begin{array}{l  l} \text{even $\ell$:} & \begin{array}{l} (\z_1 \z_2 \z_3 \ldots \z_\ell) - \tLb{2} (\z_3 \z_4 \ldots \z_\ell) - \z_1 \tLb{3} (\z_4 \ldots \z_\ell) + \tLb{2} \tLb{4} (\z_5 \ldots \z_\ell)  + \ldots  \\ ~~~
+ (\tLb{2}\tLb{4} \tLb{6} \ldots \tLb{\ell-2}) \z_{\ell-1} \z_\ell + (\tLb{2}\tLb{4} \tLb{6} \ldots \tLb{\ell-2} \tLb{\ell}), \end{array} \vspace{0.2cm}   \\
\text{odd $\ell$:} & \begin{array}{l}(\z_1 \z_2 \z_3 \ldots \z_\ell) - \tLb{2} (\z_3 \z_4 \ldots \z_\ell) - \z_1 \tLb{3} (\z_4 \ldots \z_\ell) + \tLb{2} \tLb{4} (\z_5 \ldots \z_\ell)  + \ldots  \\~~~
 + \z_1 (\tLb{3}\tLb{5} \tLb{7} \ldots \tLb{\ell} ) + \ldots + (\tLb{2}\tLb{4} \tLb{6} \ldots \tLb{\ell-1} \z_{\ell}) ; \end{array} \end{array} \right.
\end{align}
\begin{align}
(\tilde{\Pi}_{1}^{\arr} Z) &= \left\{ \begin{array}{l  l} \text{even $\arr$:} & \begin{array}{l} (Z_1 Z_2 Z_3 \ldots Z_\arr) - \Lb{2} (Z_3 Z_4 \ldots Z_\arr) - Z_1 \Lb{3} (Z_4 \ldots Z_\arr) + \Lb{2} \Lb{4} (Z_5 \ldots Z_\arr)  + \ldots \\ ~~~
+ (\Lb{2}\Lb{4} \Lb{6} \ldots \Lb{\arr-2}) Z_{\arr-1} Z_\arr + (\Lb{2}\Lb{4} \Lb{6} \ldots \Lb{\arr-2} \Lb{\arr}), \end{array} \vspace{0.2cm}   \\
\text{odd $\arr$:} & \begin{array}{l} (Z_1 Z_2 Z_3 \ldots Z_\arr) - \Lb{2} (Z_3 Z_4 \ldots Z_\arr) - Z_1 \Lb{3} (Z_4 \ldots Z_\arr) + \Lb{2} \Lb{4} (Z_5 \ldots Z_\arr)  + \ldots \\~~~
+ Z_1 (\Lb{3}\Lb{5} \Lb{7} \ldots \Lb{\arr} ) + \ldots + (\Lb{2}\Lb{4} \Lb{6} \ldots \Lb{\arr-1} Z_{\arr}) . \end{array} \end{array} \right. \label{eq:sharr}
\end{align}
The constraint equations include:
\begin{align}
\begin{array}{c}
K_L^m J_L = \x (\tilde{\Pi}_{1}^{\ell} \z) \\
K_R^m J_R = X (\tilde{\Pi}_{1}^{\arr} Z) 
\end{array} &&
\begin{array}{c}
K_L^{m-1} J_L^3 = \y (\tilde{\Pi}_{1}^{\ell} \z) \\
K_R^{m-1} J_R^3 = Y (\tilde{\Pi}_{1}^{\arr} Z) 
\end{array} &&
\begin{array}{c}
\x \y = 0  \\
X Y = 0 .
\end{array} \label{eq:qmconst}
\end{align}
Not shown above, $X$, $Y$, $\x$, and $\y$ each carry an $SU(4)$ gauge index, which is summed over in the expressions $\x^\alpha \y_\alpha = X_\alpha Y^\alpha = 0$. 
Each term in the equations above is invariant under the $SU(N)_L \times SU(N)_R$ family symmetry. 
Combinatoric coefficients have been suppressed for clarity.

\begin{table}[t]
\centering
\begin{tabular}{| c | c | c c | c |} \hline
   	 &$SU(4)$&$SU(N)_L$&$SU(N)_R$	&$\upq$	\\ \hline
$\x_1$ 	&	\yFb \Tstrut	&		& 		&	0		\\ 
$\y_1$	&	\yF	& 		&		&	0		\\
$\z_i$	&	1	& 		&		&	0		\\ \hline
$J_L$	&	\yFb \Tstrut	& 	\yF	&		&	0		\\
$K_L$	&	1	& 	\yA	&	\Bstrut	&	0 		\\ \hline
$X_1$	&	\yF	& 		&		&	1		\\
$Y_1$	&	\yFb	& 		&		&	$-1$		\\
$Z_i$	&	1	& 		&		&	0		\\ \hline
$J_R$	&	\yF	& 		&	\yF	&	$\frac{2-N}{N}$	\Tstrut	\\
$K_R$	&	1	& 		&	\yA \Bstrut	&	$4/N$ 		\\ \hline
\end{tabular}
\caption{Operators describing infrared degrees of freedom in the confined phase of $SU(N)^\ell \times SU(N)^\arr$, and their
transformation properties under the approximate $SU(N)_L \times SU(N)_R \times \upq$ flavor symmetries.}  
\label{table:IRtheory}
\end{table}

The analysis is simplified by
introducing spurion superfields $X_{i>1}$, $Y_{i>1}$, $\x_{i>1}$ and $\y_{i>1}$, 
such that the constraints between operators follow directly from the dynamically generated superpotential $W_d = W_L + W_R$, where
\begin{align}
W_L &= \frac{\x_1 \y_1 \z_1 - \x_1 \y_2 - \y_1 \x_2}{\tLb{1}}  + \sum_{i=2}^{\ell-1} \frac{\x_i \y_i \z_i - \x_i \y_{i+1} - \y_i \x_{i+1} }{\tLb{1} \tLb{2} \ldots \tLb{i} } 
+ \frac{\x_\ell \y_\ell \z_\ell - \x_\ell K_L^{m-1} J_L^3 - \y_\ell K_L^m J_L }{\tLb{1} \tLb{2} \ldots \tLb{\ell}} \\
W_R &= \frac{X_1 Y_1 Z_1 - X_1 Y_2 - Y_1 X_2}{\Lb{1}}  + \sum_{i=2}^{\arr-1} \frac{X_i Y_i Z_i - X_i Y_{i+1} - Y_i X_{i+1} }{\Lb{1} \Lb{2} \ldots \Lb{i} } 
+ \frac{X_\arr Y_\arr Z_\arr - X_\arr K_R^{m-1} J_R^3 - Y_\arr K_R^m J_R }{\Lb{1} \Lb{2} \ldots \Lb{\arr}} .
\end{align}
Each of the fields $\{X_{i>1},Y_{i>1},\x_{i>1},\y_{i>1} \}$ is a redundant operator: that is, 
the equations of motion determine the low-energy behavior of each superfield exactly, leaving no independent degrees of freedom. 
For example, the constraint $\partial W_d/ \partial X_i = 0$ determines the value of $Y_{i+1}$:
\begin{align}
Y_2 = Y_1 Z_1 &,& Y_3 =Y_1 (Z_1 Z_2 - \Lb{2}) &,& Y_{i+1} = Y_i Z_i - \Lb{i} Y_{i-1} = Y_1 (\tilde{\Pi}_1^i Z).
\end{align}

After confinement, the tree-level superpotential \eqref{eq:wtree} leads to
\begin{equation}
\wtree \rightarrow
\frac{(K_L)_{i_1 i_2} }{M_A^{2\ell-2}} + \frac{Z_1}{M_B^{N-3}} + \frac{\z_1}{M_C^{N-3}} + \frac{X_1^\alpha Y_1^\alpha}{M_R^{N-1}} 
+ \frac{\x_1^\alpha \y_1^\alpha}{M_r^{N-1}},
\end{equation}
where the indices $i$ and $\alpha$ refer to $SU(N)_L$ and $SU(4)$, respectively.
In the discussion that follows, we assume that $M_B$ is several orders of magnitude below $\mpl$, and that $M_B \lesssim M_{A,C,R,r} \lesssim \mpl$.

\subsection{Symmetry Breaking} \label{sec:ssb}

Each term in \wtree\ is introduced to break an undesired global symmetry: 
however, the  $Z_1$ and $\z_1$ tadpoles induced by $\wtree$ also have a significant effect on the vacuum structure. 
Added to the full superpotential,
\begin{equation}
W = \wtree + W_L + W_R,
\end{equation}
the $Z_1$ and $\z_1$ tadpole terms in \wtree\ shift the moduli space away from the origin: specifically, their equations of motion cause $\ev{X_1 Y_1}$ and $\ev{\x_1 \y_1}$ to be nonzero. In this section we consider the case $\ev{X_1 Y_1} \gg \ev{\x_1 \y_1}$ and show that $SU(4)\times \upq$ is spontaneously broken to 
$SU(3)_c$. 

It is convenient to normalize the infrared operators by appropriate factors of $\Lambda_i$ so as to give them canonical mass dimension $+1$:
\begin{align}
\tJ_L \equiv \frac{J_L}{\Lambda_L^\ell} &,&
\tK_L \equiv \frac{K_L}{(\Lambda_L^\ell)^2} &,&
\tx \equiv \frac{\x_1}{\tL_1^m} &,&
\ty \equiv \frac{\y_1}{\tL_1^{m+1}} &,&
\tz_i \equiv \frac{\z_i}{\tL_i^{N-1}} \\
\tJ_R \equiv \frac{J_R}{\Lambda_R^\arr} &,&
\tK_R \equiv \frac{K_R}{(\Lambda_R^\arr)^2} &,&
\tX \equiv \frac{X_1}{\Lambda_1^m} &,&
\tY \equiv \frac{Y_1}{\Lambda_1^{m+1}} &,&
\tZ_i \equiv \frac{Z_i}{\Lambda_i^{N-1}}
\end{align}
where
\begin{align}
\Lambda_L^\ell \equiv (\tL_1 \tL_2 \ldots \tL_\ell) &,&
\Lambda_R^\arr \equiv (\Lambda_1 \Lambda_2 \ldots \Lambda_\arr).
\end{align}
In terms of these operators, the tree-level superpotential \eqref{eq:wtree} becomes
\begin{eqnarray}
\wtree & \rightarrow & \Lambda_L^2 \left(\frac{\Lambda_L}{M_A} \right)^{2\ell-2} (\tK_L)_{i_1 i_2} + \Lambda_1^2 \left( \frac{\Lambda_1}{M_B} \right)^{N-3} \tZ_1 
+  \tL_1^2 \left( \frac{\tL_1}{M_C} \right)^{N-3} \tz_1 \nonumber\\&&~~~~+  \Lambda_1 \left( \frac{\Lambda_1}{M_R}\right)^{N-1} \tX \tY 
+ \tL_1 \left( \frac{\tL_1}{M_r}\right)^{N-1} \tx \ty,
\end{eqnarray} 
and the dynamically generated superpotential includes the leading terms
\begin{equation}
W_L + W_R = \tx \ty \tz_1 + \tX \tY \tZ_1 - \frac{x_1 y_2 + y_1 x_2}{\tLb{1}} - \frac{X_1 Y_2 + Y_1 X_2}{\Lb{1}} + \ldots
\end{equation}

The equation of motion $\partial W/\partial \tZ_1=0$ enforces:
\begin{equation}
\tX_\alpha \tY^\alpha = -\frac{\Lambda_1^{N-1}}{M_B^{N-3}} \equiv \sigma^2. \label{eq:mbsigma}
\end{equation}
By performing an $SU(4)$ gauge transformation, the nonzero expectation values can be rotated into the $\alpha=4$ component such that
\begin{align}
\ev{\tX}_{(4)} = \beta \sigma, && \ev{\tY}_{(4)} = \frac{1}{\beta} \sigma, && \ev{\tX}_{\alpha=1,2,3} = \ev{\tY}_{\alpha=1,2,3} = 0,
\end{align}
where $\beta$ parametrizes a flat direction of the degenerate vacua,
which is likely to be lifted in a 
particular model of \susy\ breaking; we treat it as a free parameter.
An $SU(3)_c$ subgroup of $SU(4)$ remains as an infrared symmetry, and the other $15-8=7$ generators of $SU(4)$ are broken. 
Through the super-Higgs mechanism, 7 of the 8 would-be \ngb s are eaten by the $SU(4)$ superfields 
to make them massive, and a single \ngb\ remains massless. The matter fields decompose into irreducible representations of $SU(3)_c$ as follows:
\begin{align}
\begin{array}{rcl} \yF &\longrightarrow& \yF \oplus {\yI}, \\
\tX_{\alpha'} &\longrightarrow& \tX_\alpha \oplus \tX_{(4)}, \end{array}
&&
\begin{array}{rcl} \yFb &\longrightarrow& \yFb \oplus {\yI}, \\
\tY_{\alpha'} &\longrightarrow& \tY_\alpha \oplus \tY_{(4)}, \end{array}
&&
\begin{array}{rcl} \yAd &\longrightarrow& \yAd \oplus \yF \oplus \yFb \oplus \yI, \\
\lambda_a &\longrightarrow& \lambda_a' \oplus \lambda^+ \oplus \lambda^- \oplus \lambda^0. \end{array}
\end{align}
A combination of the superfields $\tX_{\alpha=1,2,3}$ and $\tY_{\alpha=1,2,3}$ are eaten by the massive $\lambda^{\pm}$ vector supermultiplets. 
Another linear combination of $\tX$ and $\tY$ is eaten by the diagonal $T^{15}$ generator of $SU(4)$, leaving exactly one massless superfield to play the role of the axion.

We introduce the real scalar fields $\phi_1$, $\phi_2$, $a$ and $\eta$ to describe the bosonic degrees of freedom:
\begin{align} \begin{array}{rcl}
\tX_{(4)} &=& \left( \frac{\phi_1}{\sqrt2} + \ev{\tX_{(4)}} \right)\exp\left[ \frac{i}{f_a} \left( a + \alpha \eta \right) \right] \\
\tY_{(4)} &=& \left( \frac{\phi_2}{\sqrt2} + \ev{\tY_{(4)}} \right)\exp\left[ \frac{i}{f_a} \left( -a + \frac{1}{\alpha} \eta \right) \right],
\end{array} \label{eq:qcdaxion} \end{align}
where $f_a$ is the axion decay constant, and $\alpha$ is a constant determined by requiring canonical normalization of the scalar kinetic terms. It is convenient to define $v_{1,2}$ such that
\begin{align}
v_1 = \sqrt2 \abs{\ev{\tX_{(4)}}} = \sqrt2 \abs{ \beta\sigma }&&
v_2 = \sqrt2 \abs{\ev{\tY_{(4)}}} = \sqrt2 \abs{ \frac{\sigma }{\beta}},
\end{align}
so that normalization of the scalar fields requires
\begin{align}
f_a^2 = v_1^2 + v_2^2 &,&
\alpha = \frac{v_2}{v_1}.   \label{eq:fav1v2}
\end{align}
In the discussion above we assume that $\tX$ and $\tY$ are the only \upq-charged fields with nonzero expectation values. 
This is not necessarily true: for example, $\ev{K_R}$ may acquire an expectation value without breaking $SU(3)_c$. 
In the limit where $\ev{K_R} \ll \sigma$ its contribution to the axion potential is vanishingly small, and the physics remains approximately as discussed here.
For completeness, in Appendix~\ref{sec:generalkinetic} we derive the composition of the physical axion in the more general $\ev{K_R}\neq0$ case.

To preserve $SU(3)_c$ in the vacuum, the QCD-charged components of the scalars $\tx$, $\ty$, $\tJ_L$ and $\tJ_R$ must not acquire expectation values,
which places mild constraints on the unspecified nature of \susy-breaking.
Nonzero VEVs for the $i=4$ components of the scalar fields are permitted.


\subsection{Gravitational Corrections} \label{sec:grav}

Non-perturbative gravity produces \upq-violation, which at low energies are described by local gauge invariant operators in an effective superpotential.
The leading (in $1/\mpl$) terms are:
\begin{equation}
W_g = \rho_1 \frac{(\p_\ell \p_{\ell-1} \ldots \p_1 \q Q \P_1 \P_2 \ldots \P_\arr)}{\mpl^{\ell+\arr-1}} + \rho_2 \frac{(\p_\ell \p_{\ell-1} \ldots \p_1 \q )(A^m Q)}{\mpl^{\ell+m-1}}
+ \rho_3 \frac{(\barA^m \q)(A^m Q) }{\mpl^{2m-1}} + \rho_4 \frac{(A  \P_1^2 \P_2^2 \ldots \P_\arr^2)}{\mpl^{2\arr-2}},
\label{eq:wg}
\end{equation}
with coefficients $\rho_i$ which encode the details of the unknown quantum gravitational physics.
Naive power counting would argue for $\rho_i \sim \mathcal O(1)$, whereas computations based on wormhole configurations
or stringy realizations of quantum gravity
favor $\rho_i \sim \mathcal O\left( \exp \left[ -S_{\rm wh} \right] \right)$ with $S_{\rm wh} \sim \mpl / f_a$.
To capture the range of possibilities, we will consider a range of $\rho_i$ (all taken to have roughly equal magnitudes) in our analysis below.

After confinement, $W_g$ maps on to:
\begin{equation}
W_g \rightarrow \rho_1 \frac{\Lambda_L^\ell \Lambda_R^\arr}{\mpl^{\ell+\arr-1}} (\tJ_L \tJ_R) + \rho_2 \frac{\Lambda_L^\ell \Lambda_1^m}{\mpl^{\ell+m-1}} (\tJ_L \tX) 
+ \rho_3 \frac{\tL_1^m \Lambda_1^m}{\mpl^{2m-1}} (\tx \tX) + \rho_4 \frac{(\Lambda_R^\arr)^2}{\mpl^{2\arr-2}} (\tK_R)_{j_1 j_2}, \label{eq:wgconf}
\end{equation}
where the index $j$ refers to the $SU(N)_R$ family symmetry. 

There are two types of tree-level corrections to the axion potential.  In the supersymmetric limit,
the equations of motion from $\wtree+W_d+W_g$ produce operators in the Lagrangian of the form
\begin{equation}
\mathcal L_g \sim \left(\prod_{i,j} \phi_i \phi_j^\star\right)\left( \Phi + \Phi^\star\right),
\end{equation}
where $\Phi$ has non-zero \upq\ charge (and thus some of its phase is part of the axion), 
and $\phi_i$ and $\phi_j^\star$ are scalar fields as determined by the equations of motion. 
Replacing the fields with their expectation values, $\mathcal L_g$ corrects the axion potential by:
\begin{equation}
\delta V[a] \sim \left(\prod_{i,j} \ev{\phi_i} \ev{ \phi_j^\star}\right) \ev{\Phi} \cos\left( \frac{q_\Phi a}{f_a} + \theta_0 \right).
\end{equation}
Clearly this type of correction is only operative if all of the relevant fields $\phi_{i,j}$ 
have non-zero expectation values.
 
The second type of tree-level correction arises once \susy\ is broken, and the low energy Lagrangian contains
$A$-terms of the form
\begin{equation}
\mathcal L_g \sim m_s W_g + h.c.
\end{equation}
(where $W_g$ should be understood to have its super-fields replaced by their scalar components, and there is a separate \susy-breaking coefficient of
${\mathcal O}(m_s)$ for each term in $W_g$).
In the cases where the necessary scalar fields have zero expectation values, 
these terms can still correct the axion potential at loop level.

As can be seen from \eqref{eq:qmconst}, the moduli space includes vacua with $\ev{K_{R}} = \ev{J_{R}} = 0$.
These flat directions are lifted by \susy-breaking, and thus model-dependent.
Rather than getting bogged down in the details of a specific model, we make
the pessimistic assumption that the resulting expectation values are large:
\begin{equation}
 \ev{\tJ^j_{(4)}}, \ev{\tK^{j_1 j_2}} \sim \mathcal O(m_s).
\end{equation}
This assumption 
additionally simplifies the analysis in that for such large expectation values, the tree-level corrections to the axion potential are expected to
dominate over any of the loop level corrections.

Generically, the leading contributions to the axion potential are expected to arise
from \susy-breaking rather than from the equations of motion. This is because the equations of motion from $W_d$ 
involve high-dimensional operators, which are only important at tree level if all of the participating fields have relatively large expectation values. For example,
\begin{eqnarray}
\abs{ \frac{\partial W}{\partial \tJ_R} }^2 &=& \abs{ \frac{\Lambda_L^\ell \Lambda_R^\arr}{\mpl^{\ell+\arr-1} } (\tJ_L) 
- \frac{  (\tX_k \tJ_R^{2}) \tK_R^{m-1} }{\Lambda_\arr^{m}} 
- \frac{(\tY_k ) \tK_R^{m}  }{\Lambda_\arr^{m-1}} }^2 \end{eqnarray} 
reduces to 
\begin{eqnarray}
\mathcal L_g  &\sim& 
  \left( \frac{\Lambda_L^\ell \Lambda_R^\arr}{\mpl^{\ell+\arr-1} } \frac{\ev{\tK_R^m}}{\Lambda_\arr^{m-1}} \right)  \ev{\tJ_L^\star} \tY_k + h.c. \label{eq:lgsusyex}
\end{eqnarray} 
In the product $\ev{\tK_R^m}$, the $SU(N)_R$ indices are contracted antisymmetrically. If some of the expectation values are close to zero, 
the entire product vanishes.
Only in the case where $\ev{\tK}$ and $\ev{\tJ}$ are comparable to $\Lambda_\arr$ does \eqref{eq:lgsusyex} contribute significantly.

\paragraph{Quality Factors:}

The \susy-breaking $A$-term corresponding to the $\rho_1$ term in $W_g$ is
\begin{equation}
\mathcal L_g \sim m_s  \rho_1  \left(\frac{\Lambda_L^\ell \Lambda_R^\arr}{\mpl^{\ell+\arr-1}} \right) (\tJ_L)_i^\alpha (\tJ_R)^\alpha_j + h.c.,
\end{equation}
where the indices $i$ and $j$ correspond to the $SU(N)_L \times SU(N)_R$ global symmetry. 
As $\tJ_R$ is charged under \upq\, $\ev{\tJ_L \tJ_R} \neq 0$ shifts the axion potential by
\begin{equation}
\delta V[a] \sim \rho_1 m_s \left(\frac{\Lambda_L^\ell \Lambda_R^\arr}{\mpl^{\ell+\arr-1}} \right) \abs{\ev{\tJ_L} \ev{\tJ_R}} \cos\left( q_J \frac{a}{f_a} + \theta_0 \right),
\end{equation}
with $q_J = \frac{2-N}{N} = \mathcal O(1)$.
From \eqref{eq:quality}, consistency with $\abs{\bar\theta}< 10^{-10}$ requires  
\begin{equation}
\rho_1 ~\frac{m_s \mpl \abs{\ev{\tJ_L} \ev{\tJ_R}} }{\left(10^{12} \, \gev\right)^4} \left(\frac{\Lambda_L^\ell \Lambda_R^\arr}{\mpl^{\ell+\arr}} \right) < 10^{-62}. \label{eq:q1}
\end{equation}

A limit on $\arr$ is set by the $\rho_4$ term:
\begin{equation}
\delta V[a] \sim  \rho_4 m_s \frac{\Lambda_R^{2r}}{\mpl^{2r-2}} \abs{\ev{(\tK_R)_{j_1 j_2}}} \cos\left( q_K \frac{a}{f_a} + \theta_0 \right),
\end{equation}
where $q_K = 4/N $. Ignoring the $\mathcal O(1)$ number $q_K$, 
\begin{equation}
\rho_4 ~\frac{m_s \mpl^2 \abs{\ev{\tK_R}} }{\left(10^{12} \, \gev\right)^4} \left(\frac{\Lambda_R}{\mpl} \right)^{2\arr} < 10^{-62}. \label{eq:q4}
\end{equation}
From the $\rho_3$ term
\begin{equation}
\delta V[a] \sim m_s \rho_3 \frac{\tL_1^m \Lambda_1^m}{\mpl^{2m-1}} \abs{\ev{\tx_{(4)}} \ev{\tX_{(4)}} } \cos\left(\frac{a}{f_a} + \theta_0\right),
\end{equation}
we find a constraint on $N=2m+1$:
\begin{equation}
\rho_3 ~\frac{m_s \mpl \ev{\tx_{(4)}} \ev{\tX_{(4)}} }{\left(10^{12} \, \gev\right)^4} \left(\frac{\tL_1}{\mpl}\right)^m  \left(\frac{\Lambda_1}{\mpl}\right)^m < 10^{-62}. \label{eq:q3}
\end{equation}
Finally, the $\rho_2$ term sets an additional constraint on $\ell$ and $N$:
\begin{equation}
\delta V[a] \sim m_s \rho_2 \frac{\Lambda_L^\ell \Lambda_1^m}{\mpl^{\ell+m-1}} \abs{ \ev{J_L^{(4)}} \ev{\tX_{(4)}} } \cos\left(\frac{a}{f_a} + \theta_0 \right), 
\end{equation}
\begin{equation}
\rho_2 ~\frac{m_s \mpl \ev{J_L} \ev{\tX_{(4)}} }{\left(10^{12} \, \gev\right)^4} \left(\frac{\Lambda_L}{\mpl} \right)^\ell \left(\frac{\Lambda_1}{\mpl} \right)^m < 10^{-62} . \label{eq:q2}
\end{equation}
As long as $\beta$ is neither very large nor very small,
Eqs.~(\ref{eq:q1}), (\ref{eq:q4}), (\ref{eq:q3}) and~(\ref{eq:q2}) provide the most restrictive constraints on $m$, $\ell$ and $\arr$.
A wide range of values is allowed for each of the parameters, as we discuss in more detail below.


\subsection{Benchmark Models:} 


\begin{table}[t] \centering
\begin{tabular}{| c | c | } \hline
{\bf B1}		&  (\gev) \\ \hline
$f_a$		& $10^{17}$ \\
$\Lambda_1$	& $10^{17}$ \\
$\Lambda_{i>1}$& $10^{15}$ \\
$\tL_{i}$	& $10^{15}$ \\ 
$m_s$		& $10^{6}$ \\ \hline
\end{tabular}
\begin{tabular}{| c | c | } \hline
{\bf B2}		&  (\gev) \\ \hline
$f_a$		& $10^{12}$ \\
$\Lambda_1$	& $10^{12}$ \\
$\Lambda_{i>1}$& $10^{9}$ \\
$\tL_{i}$	& $10^{9}$ \\ 
$m_s$		& $10^{4}$ \\ \hline
\end{tabular}
\begin{tabular}{| c | c | } \hline
{\bf B3}		&  (\gev) \\ \hline
$f_a$		& $10^{9}$ \\
$\Lambda_1$	& $10^{9}$ \\
$\Lambda_{i>1}$& $10^{4}$ \\
$\tL_{i}$	& $10^{4}$ \\ 
$m_s$		& $10^{4}$ \\  \hline
\end{tabular}
\caption{Three benchmark points in the parameter space of $\Lambda_i$ and $\tL_i$. With the exception of $\ev{\tX}$ and $\ev{\tY}$, the expectation values of the $SU(3)_c$ singlet fields are taken to be $\mathcal O(m_s)$.}
\label{table:bench}
\end{table}

In this section we consider the quality of the axion potential in three particular models, with $f_a=10^{17} \, \gev$,  $f_a=10^{12} \, \gev$ and $f_a=10^{9} \, \gev$. For simplicity, we take $\Lambda_1 \sim M_B \sim f_a$ and $\Lambda_{i\neq 1} \sim \tL_i$ for each model, and we allow all QCD singlet scalar fields to acquire $\mathcal O(m_s)$ expectation values. Choices for each of these scales are shown in Table~\ref{table:bench}.

Model {\bf B1} is particularly susceptible to gravitational disruptions, as the scales $\Lambda_{i}$ and $\tL_i$ are taken to be relatively close to the Planck scale $\mpl\sim10^{19} \, \gev$. In this model even exponential suppression of the constants $\rho_i \sim \exp (-\mpl/f_a) \sim 10^{-44}$ cannot account for the high quality of the axion potential, and large values of $N$, $\ell$ and $\arr$ are required. Models {\bf B2} and {\bf B3} have values of $f_a \lesssim 10^{12}\, \gev$ consistent with the axion dark matter hypothesis; with its smaller values of $\Lambda_i$ and $\tL_i$, model {\bf B3} is more adept at suppressing gravitational corrections.

\begin{figure}
\centering
\includegraphics[scale=1.0]{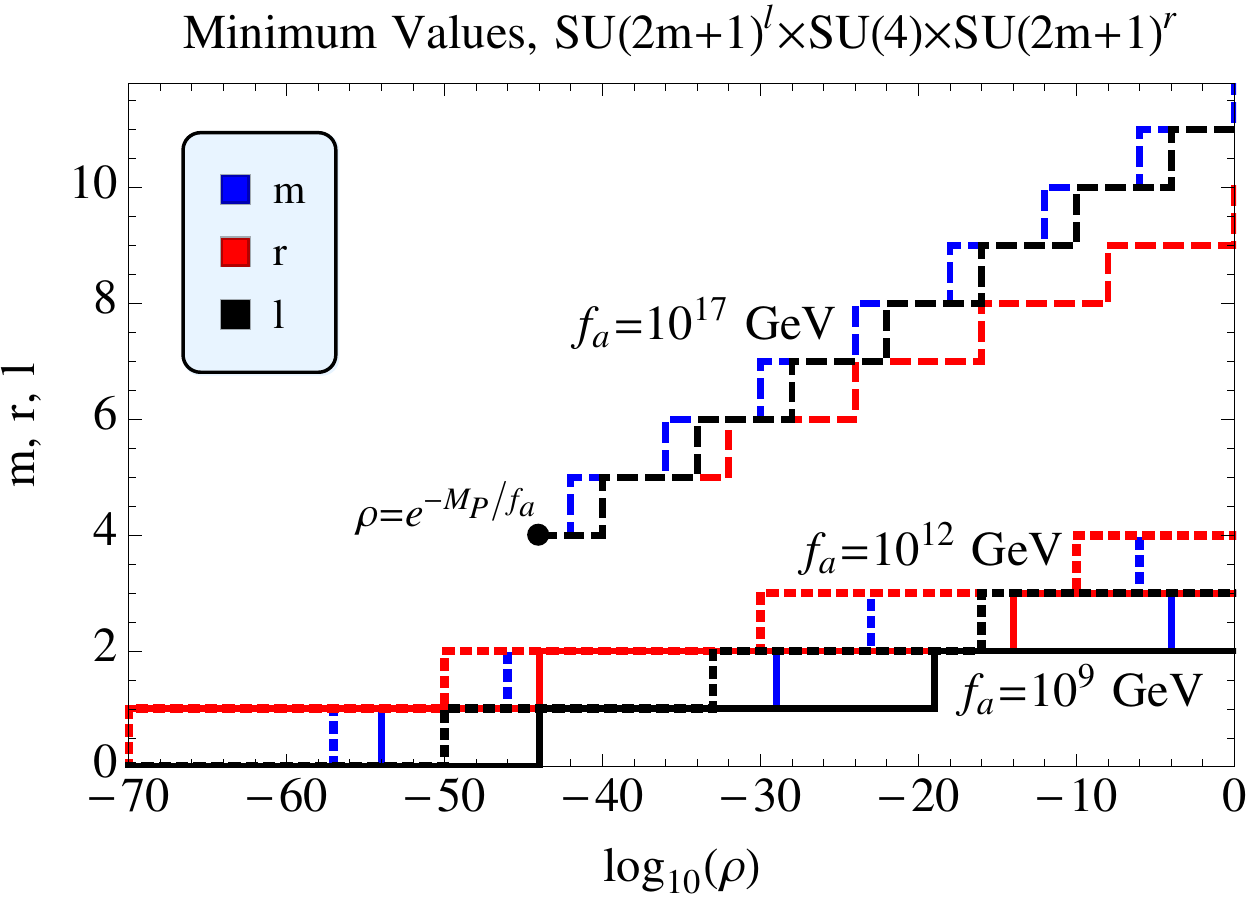}
\caption{Minimum values for $m$, $\ell$ and $\arr$ consistent with $\abs{\bar\theta} < 10^{-10}$ are shown as a function of $\rho_{1\ldots 4}$. For the first benchmark model with $f_a = 10^{17}\, \gev$, we show only values of $\rho \gtrsim \exp(-\mpl/f_a) \approx 10^{-43.4}$. The $f_a = 10^{12}\, \gev$ and $f_a = 10^{9}\, \gev$ models are depicted using dotted and solid lines, respectively.}
\label{fig:qualityplot}
\end{figure}

In Figure~\ref{fig:qualityplot} we show minimum values for $m\equiv\frac{N-1}{2}$, $\ell$, and $\arr$ consistent with $\abs{\bar\theta}<10^{-10}$ for the $SU(N)^\ell \times SU(4) \times SU(N)^\arr$ composite axion, as a function of the parameters $\rho_i$. A wide range is shown for $\rho$, to accommodate both exponentially suppressed and $\mathcal O(1)$ values. In the $\rho_i = \mathcal O(1)$ limit, the minimal gauge groups for the three benchmark models are:
\begin{align}
\begin{array}{r l}
\text{\bf B1:} ~&~ SU(23)^{11} \times SU(4) \times SU(23)^{9} \\
\text{\bf B2:} ~&~ SU(9)^3 \times SU(4) \times SU(9)^4 \\
\text{\bf B3:} ~&~ SU(7)^2 \times SU(4) \times SU(7)^3 .
\end{array} \label{eq:b123rho1}
\end{align}
Naturally, if after \susy\ breaking the scalar fields $\tJ_{L,R}$, $\tx$, $\ty$, and $\tK_R$ do not acquire expectation values, then the \upq\ violation induced by $W_g$ affects the axion potential only at loop level, and smaller values for $N$, $\ell$ and $\arr$ are permitted.  
In the limit where $\rho$ is exponentially suppressed, $\abs{\bar\theta}<10^{-10}$ no longer constrains $m$, $\ell$ or $\arr$. Although Eqs.~(\ref{eq:q1}), (\ref{eq:q4}), (\ref{eq:q3}) and~(\ref{eq:q2}) are valid only for $m \geq2$, $\arr \geq 1$ and $\ell \geq 0$, smaller values for $m$ and $\arr$ are shown in Figure~\ref{fig:qualityplot} to 
indicate where $\rho$ is small enough that compositeness is no longer necessary.

\section{Dynamically Generated \wtree} \label{sec:dynsp}

As described in Section~\ref{sec:main}, the $SU(N)^\ell \times SU(4) \times SU(N)^\arr$ composite accidental axion has a high-quality scalar potential and 
most of the important scales are derived from the confining dynamics, with the exception of $M_B$ in the tree-level superpotential. 
This is a relatively minor shortcoming: $f_a$ is determined by the relationship between $M_B$, $\Lambda_1$, and $\beta^2 = \ev{\tX}/\ev{\tY}$,
\begin{equation}
f_a^2 = 2 \abs{\frac{\Lambda_1^{N-1} }{M_B^{N-3}} \left(\beta^2 + \frac{1}{\beta^2}\right) }, \label{eq:fambbeta}
\end{equation}
and the scale $M_B \ll \mpl$ is added ``by hand" in the tree-level superpotential. 
In this section we show how the $M_B$ term in \wtree\ can be 
dynamically generated by the s-confinement of an $Sp(2N-4)$ gauge group, so that all of the important mass scales are determined by strong dynamics.

A gauge theory with $2N$ quarks $\sq$ charged under $Sp(2N-4)$ in the fundamental representation
s-confines~\cite{Intriligator:1995ne} to form mesons $M_{i j}=  \epsilon_{ab} \sq^a_i \sq^b_j$, with the superpotential
\begin{equation}
W_d = \frac{\Pf M}{\Lambda_0^{2N-1}}.
\end{equation} 
We break the $SU(2N)$ flavor symmetry by gauging its $SU(N)_1 \times SU(N)_2 = G_1 \times G_2$ subgroup:
\begin{align}
{\yF} \longrightarrow (\yF, \yI) \oplus(\yI, \yF) &&
\sq^a_i \longrightarrow (\sq_1)_\alpha^a \oplus (\sq_2)^a_\beta,
\end{align}
where $\alpha$ and $\beta$ correspond respectively to the $SU(N)_1$ and $SU(N)_2$ gauge indices. The meson $M \sim \yA$ decomposes into irreducible representations of $G_1 \times G_2$:
\begin{align}
\tM_1^{\alpha_1 \alpha_2} = \frac{(\sq_1)_a^{\alpha_1} (\sq_1)_b^{\alpha_2} \epsilon_{ab}}{\Lambda_0} &,&
\barQ_1^{\alpha\beta} = \frac{(\sq_1)_a^{\alpha} (\sq_2)_b^{\beta} \epsilon_{ab}}{\Lambda_0}  &,&
\tM_2^{\beta_1 \beta_2} = \frac{(\sq_2)_a^{\beta_1} (\sq_2)_b^{\beta_2} \epsilon_{ab}}{\Lambda_0}  ,
\end{align}
where $\Lambda_0$ is the confinement scale of $Sp(2N-4)$.
In terms of these operators the dynamically generated superpotential is
\begin{equation}
W_d = \frac{\Pf(\sq^2)}{\Lambda_0^{2N-3} } = \frac{(\Lambda_0)^{N}}{\Lambda_0^{2N-3}} \left[ \tM_1^m \barQ_1 \tM_2^m + \tM_1^{m-1} \barQ_1^{3} \tM_2^{m-1} + \ldots + 
\tM_1 \barQ_1^{2m-1} \tM_2  + \barQ_1^{2m+1} \right],  \label{eq:wdsp}
\end{equation}
in the case where $N=2m+1$ is odd.
Combinatoric factors for each term in the expansion of $\Pf M$ such as $\barQ_1^N \equiv \det \P_1$ have been suppressed.

\begin{figure}[t] \centering
\includegraphics[scale=1.0]{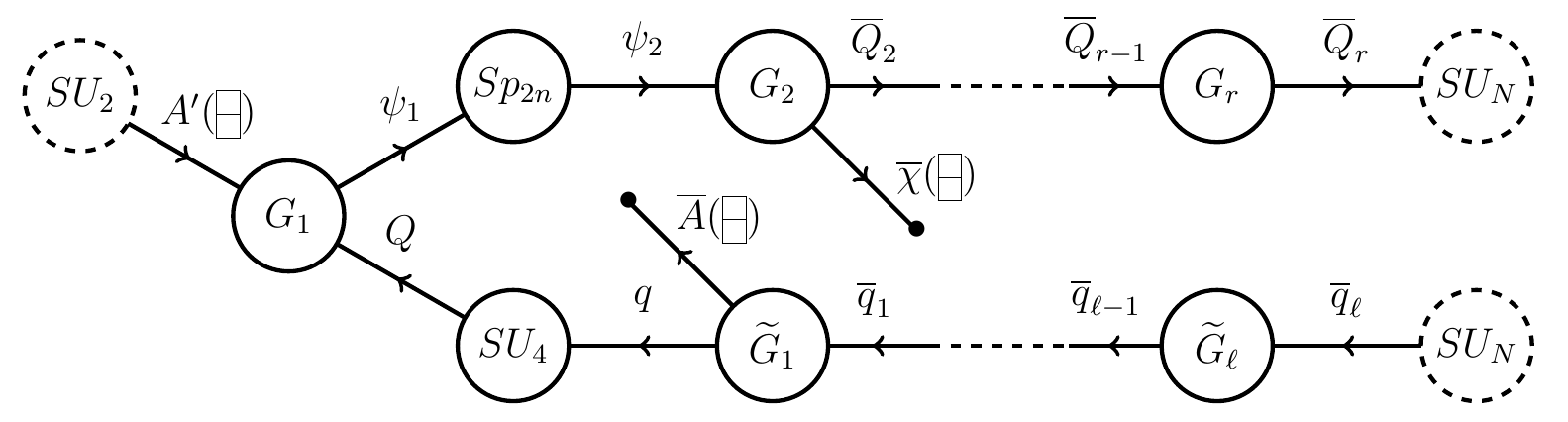}
\caption{The matter content of the $SU(N)^\ell\times SU(4)\times Sp(2n) \times SU(N)^\arr$ composite axion model is depicted in the moose diagram above, with $Sp_{2n} \equiv Sp(2N-4)$. The $SU(2)$ family symmetry of the $A'$ fields is broken explicitly by the tree-level superpotential \eqref{eq:wtreesp}.
 }
\label{fig:spmoose}
\end{figure}

To match this theory with the $A+4Q +N\barQ$ model, the $M_1$ and $M_2$ degrees of freedom must be removed. This is achieved by adding the following matter fields charged under $SU(N)_1 \times SU(N)_2$:
\begin{equation}
2 A' + 4 Q + \chi + N \barQ_2 = 2 {(\yA, \yI)} \oplus 4 {(\yF, \yI)} \oplus {(\yI, \yAb)} \oplus N {(\yI, \yFb)} .
\end{equation}
In the $SU(N)^\ell \times SU(4) \times SU(N)^\arr$ composite model, the $SU(4)$ and $SU(N)$ family symmetries of the $Q$ and $\barQ_2$ are gauged. The full matter content of the theory is shown in Figure~\ref{fig:spmoose}.

Gauge-invariant operators of the form $(A' \sq_1^2)$ and $(\chi \sq_2^2)$ can be added as marginal operators in a tree-level superpotential:
\begin{equation}
\wtree = \lambda_i (A'_i)^{\alpha_1\alpha_2} (\sq_1)_{\alpha_1}^{a_1} (\sq_1)_{\alpha_2}^{a_2} \epsilon_{a_1 a_2} + \lambda_0 \chi^{\beta_1 \beta_2} (\sq_2)_{\beta_1}^{a_1} (\sq_2)_{\beta_2}^{a_2} \epsilon_{a_1 a_2}, \label{eq:wtreesp}
\end{equation}
where the indices $i$, $a$, $\alpha$ and $\beta$ correspond to $SU(2)$, $Sp(2N-4)$, $SU(N)_1$ and $SU(N)_2$, respectively, and $\lambda_i$ and $\lambda_0$ are dimensionless coupling constants.
After $Sp(2N-4)$ confines, \wtree\ becomes
\begin{equation}
\wtree = \lambda_i \Lambda_0 (A'_i)^{\alpha_1\alpha_2} \tM_1^{\alpha_1 \alpha_2} + \lambda_0 \Lambda_0 \chi^{\beta_1 \beta_2} \tM_2^{\beta_1 \beta_2} .
\end{equation}
This is extremely convenient: in the limit where $\Lambda_0 \gg \Lambda_1$, the fields $M_1$, $M_2$, $\chi$, and the linear combination ``$(A'_1+ A'_2)$" all acquire large masses and decouple. 
One linear combination of $A'_1$ and $A'_2$ remains massless, which we define as $A$:
\begin{equation}
A \equiv \frac{\lambda_2 A_1 - \lambda_1 A_2}{\mathcal N},
\end{equation}
with some normalization factor $\mathcal N$.

The dynamically generated superpotential simplifies greatly when we consider the fact that $\tM_1$ and $\tM_2$ have $\mathcal O(\Lambda_0)$ masses from \wtree:
\begin{align}
\frac{\partial W}{\partial A'_i} = \lambda_i \Lambda_0 \tM_1 &,& \frac{\partial W}{\partial \chi} = \lambda_0 \Lambda_0 \tM_2. 
\end{align}
After integrating out the heavy fields, the superpotential becomes
\begin{equation}
W = \frac{\barQ_1^N}{\Lambda_0^{N-3}}.
\end{equation}
Not only is this the desired tree-level superpotential for the composite axion model, but all of the extra matter fields $A'$, $\chi$, $\tM_1$ and $\tM_2$ have decoupled, leaving only $A$ and $\barQ_1$ as infrared degrees of freedom.  
In \eqref{eq:fambbeta} $M_B$ is replaced by $\Lambda_0$, so that
\begin{equation}
f_a^2 = 2 \abs{\frac{\Lambda_1^{N-1} }{\Lambda_0^{N-3}} \left(\beta^2 + \frac{1}{\beta^2}\right) }. \label{eq:fasp}
\end{equation}
Every important scale other than \mpl\ is now determined solely by confining dynamics.

\begin{table}[t]
\centering
\begin{tabular}{| c | c | c c c  c | c |  c |} \hline
   	&$Sp(2N-4)$&$SU(N)_1$&$SU(N)_2$&$SU(N)_3$&$SU(4)$	&$SU(2)$	& \upq	\\ \hline
$\sq_1$  &\yF		&	\yFb	&			&		&		&		&	$-2/N$		\\ 
$\sq_2$&	\yF		&		&	\yF		&		&		&		&	$+2/N$		\\  \hline 
$A'$&			& 	\yA \Tstrut	&			&		&		&	\bf 2	& $4/N$		\\ 
$\chi$	&		&		&	\yAb	\Bstrut	&		&		&	\bf 1	&	$-4/N$		\\ \hline
$Q$&			& 	\yF	&			&		&	\yF	&		&$\frac{2-N}{N}$ \Tstrut		\\ 
$\barQ_2$&		&		&	\yFb		&	\yF	& 		&		&	0		\\ \hline
\end{tabular}
\caption{A subset of the matter fields in the $Sp(2N-4)$ model are shown with their Peccei-Quinn charges. All of the non-Abelian groups except for $SU(2)$ are gauged.
}  
\label{table:spcomplete}
\end{table}

The nonzero $Sp(2N-4)^2$-$U(1)_B$ anomaly breaks $U(1)_B$ explicitly, as can be seen from the $W_d$ of \eqref{eq:wdsp}. Although in principle the new fields $\chi$ and $A'$ provide two additional anomaly-free $U(1)$ symmetries, these are broken by the tree-level superpotential \eqref{eq:wtreesp}, and only the $SU(N)_L \times SU(N)_R \times U(1)_A \times U(1)_C \times U(1)_R$ global symmetry remains. Introducing
\begin{equation}
\delta \wtree = \frac{ (\barA \barq_1^2 \barq_2^2 \ldots \barq_\ell^2)}{M_A^{2\ell-2}} + \frac{(\p_1^N)}{M_C^{N-3}} + \frac{(A^{m} Q)(A^{m-1} Q^3)}{M_R^{N-1} } + \frac{(\barA^{m} \q)(\barA^{m-1} \q^3)}{M_r^{N-1} } 
\end{equation}
with $M_A \sim M_C \sim M_R \sim M_r \sim \mpl$ is sufficient to give masses to the additional \pngb s. 
In Table~\ref{table:spcomplete}, the Peccei-Quinn charges of each field is shown.

\paragraph{Axion Quality:}

Of the new superpotential terms which break \upq, the leading terms are
\begin{equation}
W_g \sim \frac{\chi^m \barQ_2 \barQ_3 \ldots \barQ_\arr}{\mpl^{m+\arr-4}}   + \sum_{p} \frac{(A_1^{m-p} A_2^p Q)(\q \barq_1 \barq_2  \ldots \barq_\ell)}{\mpl^{m+\ell-1}}
\end{equation}
As $\chi$ has a mass of $\mathcal O(\Lambda_0)$ and no expectation value, the $\chi^m$ interaction has no tree-level effect on the axion potential. The only effects are loop-induced and receive additional suppression. 

One linear combination in the $(A_1^{m-p} A_2^p Q)$ sum corresponds to the infrared operator $(A^m Q)$, which has the expectation value $\ev{X_1}$. This term is already included in the $W_g$ of \eqref{eq:wg}. Every other term in the sum includes a power of the massive combination $(\lambda_1 A_1 + \lambda_2 A_2)$, which has no expectation value, and is therefore less disruptive to the axion potential than the effects already considered in \eqref{eq:wg}.

Aside from the replacement of $M_B$ by $\Lambda_0$, the quality factors calculated in Section~\ref{sec:grav} are largely unchanged. Operators involving $\barQ_1$ are the exception: now that $\barQ_1 = \sq_1 \sq_2 /\Lambda_0$, a suppression of $\Lambda_0/\mpl$ is added to the operators involving $J_R$ and $K_R$, marginally improving Eqs.~(\ref{eq:q1}) and~(\ref{eq:q4}):
\begin{align}
\rho_1\, \frac{m_s \mpl \abs{\ev{\tJ_L} \ev{\tJ_R}} }{\left(10^{12} \, \gev\right)^4} \left(\frac{\Lambda_0}{\mpl}\right) \left(\frac{\Lambda_L^\ell \Lambda_R^\arr}{\mpl^{\ell+\arr}} \right) ~<~& 10^{-62} \label{eq:q1b} \\
\rho_4 \, \frac{m_s \mpl^2 \abs{\ev{\tK_R}} }{\left(10^{12} \, \gev\right)^4} \left(\frac{\Lambda_0}{\mpl}\right)^2 \left(\frac{\Lambda_R}{\mpl} \right)^{2\arr} ~<~& 10^{-62}. \label{eq:q4b}
\end{align}
For many values of $\rho_i$ this decreases the minimum value for $\arr$ by one, as can be seen from the three benchmark models at $\rho_i = \mathcal O(1)$:
\begin{align}
\begin{array}{r l}
\text{\bf B1:} ~&~ SU(23)^{11} \times SU(4)\times Sp(42) \times SU(23)^{9} \\
\text{\bf B2:} ~&~ SU(9)^3 \times SU(4)\times Sp(14) \times SU(9)^3 \\
\text{\bf B3:} ~&~ SU(7)^2 \times SU(4) \times Sp(10) \times SU(7)^2 .
\end{array} 
\end{align}

\paragraph{Alternate Confinement Order:}

Thus far, we have required that $\Lambda_0 > \Lambda_1$, simply because the dual of $SU(N): 2 A + 4 Q + (2N-4) \barQ$ with the tree-level superpotential $\wtree \sim A \barQ^2$ does not appear in the literature. In principle the infrared behavior of the $2A+4 Q + (2N-4) \barQ$ theory with $\wtree\neq 0$ can be determined using ``deconfinement" techniques~\cite{Berkooz:1995km} and a sequence of dualities: a similar calculation~\cite{Craig:2011wj} has been completed for $A+F Q + (N+F-4)\barQ$ with a superpotential of the form $W\sim A \barQ^2$.

Without calculating the degrees of freedom and the superpotential in the infrared dual of $SU(N): 2A + 4Q + (2N-4)\barQ$, it is not known how the scale $f_a$ is set in the dual theory.
If in the $\Lambda_0 \ll \Lambda_1$ limit $\upq$ is still broken at the scale $f_a^2 \sim \Lambda_1^{N-1} / \Lambda_0^{N-3}$, then $f_a \sim 10^{12}\, \gev$ can be achieved with much smaller values of $\Lambda_0$ and $\Lambda_1$, significantly improving the axion quality. 
We leave detailed exploration of this limit to future work.

\section{Conclusions}
\label{sec:conclusions}

In the composite axion model based on the gauge group $SU(N)^\ell \times SU(4) \times SU(N)^\arr$, a \upq\ is spontaneously broken by the vacuum expectation values of the $SU(4)$-charged hadrons $X_1 = (A^m Q)$ and $Y_1 = (A^{m-1} Q^3)$, simultaneously producing the QCD axion and breaking $SU(4)$ to $SU(3)_c$. All important scales in the axion model are generated dynamically from confinement, and are naturally small compared to the Planck scale.

By calculating the disruption to the axion potential $V[a]$ induced by Planck-scale effects, we have demonstrated that the composite model is successful at preserving the quality of the axion potential even when large expectation values are permitted for all of the \upq-charged QCD-singlet scalar fields. In realistic models incorporating \susy\ breaking with positive quadratic terms for these scalars such that no large expectation values result, the quality of the axion potential will improve significantly for any given $N$, $\ell$ and $\arr$, as the terms in $W_g$ disrupt the axion potential to a lesser degree.  It would be worthwhile to further investigate such constructions.

It is likely that the success of the $SU(N)^\ell \times SU(4) \times SU(N)^\arr$ composite axion can be replicated by embedding $SU(3)_c$ within the $SU(N)_R$ flavor symmetry of the $A+4Q + N \barQ$ model. In this case \upq\ will be more closely associated with the $U(1)_B$ flavor symmetry of Table~\ref{table:UVtheory} rather than $U(1)_A$, and the axion will be generated from a linear combination of $(\barQ_i^N)$ baryons.

Compositeness can cure the axion quality problem, and as our models demonstrate, may provide clues to the existence of interesting dynamics in the ultraviolet.




\section*{Acknowledgments}

This research was supported in part by the NSF grant PHY-1316792. 
The authors are grateful for helpful conversations with A.~Rajaraman, M.~Ratz, Y.~Shirman, and P.~Tanedo.

\appendix 

\section{Axion Quality} \label{appx:quality}

To leading order in $a$, the QCD-induced axion potential $V[a]$ has the form
\begin{equation}
V[a] = V_0 - \frac{1}{2} m_a^2 \left( a + f_a \bar\theta \right), \label{eqx:va}
\end{equation}
which is minimized when $\ev{\theta} \equiv (a/f_a + \bar\theta)$ is equal to zero. It is convenient to define the shifted field $\alpha \equiv a + f_a \bar\theta$, so that $\ev{\theta} = \ev{\alpha}/f_a$.
Explicit \upq\ violation elsewhere in the theory adds corrections to $V[a]$,
\begin{equation}
\delta V[a] = Q f_a^4 \cos \left( \kappa\left[\frac{a}{f_a} + \bar\theta \right] + \theta_0 \right), \label{eqx:dva}
\end{equation}
which for small values of $\ev{\theta}$ is approximately
\begin{equation}
\delta V[a] = Q f_a^4 \left[1 - \frac{1}{2} \left( \frac{\kappa \alpha}{f_a} \right)^2 \right] \cos \theta_0 - Qf_a^4 \left(\frac{\kappa \alpha}{f_a} \right) \sin \theta_0 .
\end{equation}
As $\theta_0$ is determined by the precise manner in which \upq\ is broken, we do not assume that it is smaller than $\mathcal O(1)$. Combining Eqs.~(\ref{eqx:va}) and~(\ref{eqx:dva}), $V[a]$ becomes
\begin{equation}
V[\alpha] = \left( V_0 + Qf_a^4 \cos\theta_0 \right) -\left( Q f_a^3 \kappa \sin\theta_0 \right) \alpha - \frac{1}{2} \left( m_a^2 + Q f_a^2 \kappa^2 \cos\theta_0 \right) \alpha^2,
\end{equation}
so that the expectation value $\ev{\alpha}$ shifts away from zero:
\begin{equation}
\ev{\alpha} = - \frac{Q f_a^3 \kappa \sin\theta_0 }{m_a^2 + Q f_a^2 \kappa^2\cos\theta_0 }.
\end{equation}
Experimental measurements of $\ev{\theta}$ set an upper bound $\ev{\alpha} < f_a \thmax$. Assuming $\thmax \kappa \ll \sin\theta_0$, the corresponding bound on $Q$ is
\begin{equation}
Q < \frac{m_a^2}{f_a^2} \frac{\thmax}{\kappa \abs{\sin\theta_0} }. \label{eqx:qbound}
\end{equation}
Using $m_a^2 = m_\pi^2 f_\pi^2/f_a^2$ and
assuming $\kappa \sin\theta_0 = \mathcal O(1)$, \eqref{eqx:qbound} implies
\begin{equation}
Q \lesssim 10^{-62}  \left(\frac{10^{12} ~\gev}{f_a} \right)^4.
\end{equation}

\section{Axion Assignment in a General Vacuum} \label{sec:generalkinetic}

Suppose there exist many fields $\Phi_i$, each with a Peccei-Quinn charge $q_i$. Let us define the charge-normalized expectation value
\begin{equation}
v_i \equiv q_i \sqrt2 \ev{\Phi_i}
\end{equation}
for each field $\Phi_i$. If there are $n$ fields with nonzero expectation values, then let us define $n -1$ fields $\eta_i$ and the axion $a$, with the following assignment:
\begin{eqnarray}
\Phi_1 &=& \left(\frac{\phi_1}{\sqrt2} + \ev{\Phi_1}\right)\exp\left[ \frac{i q_1}{f_a}( a + \alpha_1 \eta_1) \right] \\
\Phi_2 &=& \left(\frac{\phi_2}{\sqrt2} + \ev{\Phi_2}\right)\exp\left[ \frac{i q_2}{f_a}( a + \beta_1 \eta_1 + \beta_2 \eta_2) \right] \\
\Phi_3 &=& \left(\frac{\phi_3}{\sqrt2} + \ev{\Phi_3}\right)\exp\left[ \frac{i q_3}{f_a}( a + \gamma_1 \eta_1 + \gamma_2 \eta_2 + \gamma_3 \eta_3) \right] \\
&\vdots& \nonumber\\
\Phi_{n-1} &=& \left(\frac{\phi_{n-1}}{\sqrt2} + \ev{\Phi_{n-1}}\right)\exp\left[ \frac{i q_{n-1}}{f_a}( a + \alpha^{(n-1)}_1 \eta_1 + \ldots + \alpha^{(n-1)}_{n-1} \eta_{n-1} ) \right] \\
\Phi_{n} &=& \left(\frac{\phi_{n}}{\sqrt2} + \ev{\Phi_{n}}\right)\exp\left[ \frac{i q_{n}}{f_a}( a + \alpha^{(n)}_1 \eta_1 + \ldots + \alpha^{(n)}_{n-1} \eta_{n-1} ) \right] 
\end{eqnarray}
In the sequence above, the first appearance of each field $\eta_i$ is in the phase of $\Phi_i$. The field $\Phi_n$ does not introduce any new $\eta_i$ fields.

Let us define the following $(n-1)$ constants:
\begin{eqnarray}
x_1 &=& \beta_1 = \gamma_1 = \delta_1 = \ldots = \alpha^{(n-1)}_1 = \alpha^{(n)}_1 \\
x_2 &=& \gamma_2 = \delta_2 = \ldots = \alpha^{(n-1)}_2 = \alpha^{(n)}_2 \\
x_3 &=& \delta_3 = \ldots = \alpha^{(n-1)}_3 = \alpha^{(n)}_3 \\
&\vdots& \nonumber\\
x_{n-2} &=& \alpha^{(n-1)}_{n-2} = \alpha^{(n)}_{n-2} \\
x_{n-1} &=& \alpha^{(n)}_{n-1}.
\end{eqnarray}
These equalities follow from the vanishing of the kinetic cross terms, which also give the following relationships between the $x_i$ and $\{\alpha_1, \beta_2, \gamma_3, \ldots, \alpha^{(n-1)}_{n-1} \}$:
\begin{eqnarray}
0 &=& 1 + x_1 \alpha_1 \\
0 &=& 1 + x_1^2 + x_2 \beta_2 \\
0 &=& 1 + x_1^2 + x_2^2 + x_3 \gamma_3 \\
&\vdots& \nonumber\\
0 &=& 1 + x_1^2 + \ldots + x_{n-2}^2 + x_{n-1} \alpha^{(n-1)}_{n-1} .
\end{eqnarray}
Finally, we require that the kinetic terms $(\partial_\mu \eta_i)^2$ and $(\partial_\mu a)^2$ are canonically normalized. This leads to the remaining $n$ constraints:
\begin{eqnarray}
\frac{f_a^2}{v_1^2} &=& 1 + \alpha_1^2  \\
\frac{f_a^2}{v_2^2} &=& 1+ x_1^2 + \beta_2^2 \\
\frac{f_a^2}{v_3^2} &=& 1+ x_1^2 + x_2^2 + \gamma_3^2 \\
&\vdots & \nonumber\\
\frac{f_a^2}{v_{n-1}^2} &=& 1+ x_1^2 + x_2^2 + \ldots  + x_{n-2}^2 + (\alpha^{(n-1)}_{n-1})^2 \\
\frac{f_a^2}{v_{n}^2} &=& 1+ x_1^2 + x_2^2 + \ldots  + x_{n-2}^2 + x_{n-1}^2 .
\end{eqnarray}

These systems of equations have the solutions:
\begin{align}
\alpha_1^2 &= \frac{f_a^2 - v_1^2}{v_1^2} &
x_1^2 &= \frac{v_1^2}{f_a^2 - v_1^2} \\
\beta_2^2 &= \frac{f_a^2 (f_a^2 -v_1^2 - v_2^2) }{v_2^2 (f_a^2 - v_1^2) } &
x_2^2 &= \frac{v_2^2 f_a^2}{(f_a^2 - v_1^2 - v_2^2)(f_a^2  - v_1^2) } \\
\gamma_3^2 &= \frac{f_a^2 (f_a^2 -v_1^2 - v_2^2 - v_3^2) }{v_3^2 (f_a^2 - v_1^2 - v_2^2) } &
x_3^2 &= \frac{v_3^2 f_a^2}{(f_a^2 - v_1^2 - v_2^2 - v_3^2)(f_a^2  - v_1^2 - v_2^2) },
\end{align}
and so on. The general solution is
\begin{eqnarray}
(\alpha^{(i)}_i )^2 &=& \frac{f_a^2 (f_a^2 - v_1^2 - v_2^2 - \ldots - v_i^2) }{v_i^2 (f_a^2 - v_1^2 - v_2^2 - \ldots - v_{i-1}^2) } \\
x_i^2 &=& \frac{v_i^2 f_a^2}{(f_a^2 - v_1^2 - v_2^2 - \ldots - v_i^2)(f_a^2 - v_1^2 - v_2^2 - \ldots - v_{i-1}^2) },
\end{eqnarray}
for $i = 1 \ldots (n-1)$. 
Each $\alpha^{(i)}_i$ and $x_i$ must also obey
\begin{equation}
\alpha^{(i)}_i x_i < 0,
\end{equation}
but the signs of $\alpha^{(i)}$ and $x_i$ are otherwise arbitrary.

Finally, the axion decay constant is: 
\begin{eqnarray}
f_a^2 &=& v_1^2 + v_2^2 + \ldots + v_{n-1}^2 + v_n^2 .
\end{eqnarray}


\bibliography{compaxion}

\end{document}